# Large area Silicon tracking: New perspectives

Aurore Savoy-Navarro

CNRS-IN2P3/University Paris-Diderot
Paris-75205 – France

The successful running of the large area Silicon trackers of ATLAS and CMS at LHC, and the ongoing R&D for the upgrade of these tracking systems, in various stages, over this decade, are a full proof of this technology and of its still impressive potential. The Linear Collider project is waiting for the possible discovery of a light Higgs at LHC maybe by end of 2012. These facts opened a new phase for the R&D on Silicon tracking for the Linear Collider, with enhanced synergy with LHC, Astrophysics and other HEP experiments, thus leading to new perspectives and alternatives.

## 1 Introduction

Over the last two years, the successful running of the ATLAS, CMS and LHCb experiments has been quite impressive [1]. The challenging all-Silicon tracking system built by CMS achieved high performances. As an example, Figure 1 shows a detailed reconstructed view of an event, with a very high charged track density and harsh environmental conditions. Likewise the Silicon tracking components that covers the inner part of the ATLAS tracking system or the LHCb Silicon tracking are other real life full proofs of the performances of this tracking technology.

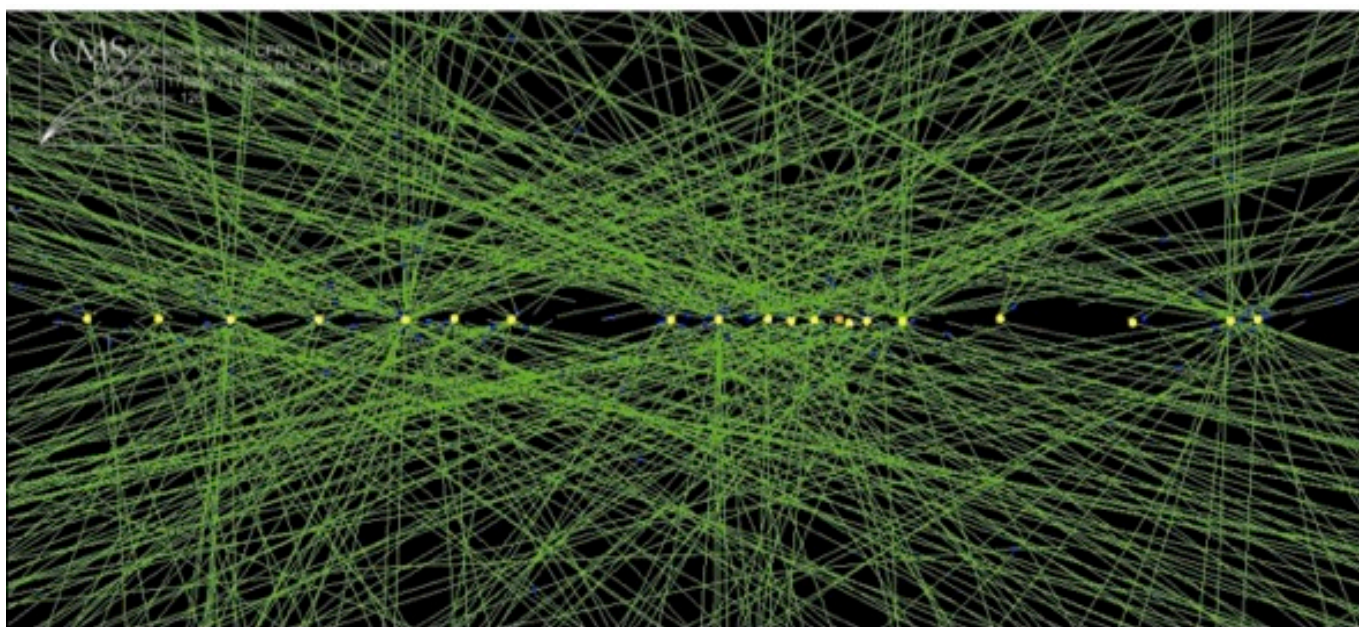

Figure 1: Pile up event as visualized by the CMS Silicon tracker

In addition, these Silicon tracking systems are an invaluable source of information on how this technology can and must be further improved.
Besides, the decision on the Linear Collider (LC) project is pending, waiting for a possible discovery of a light Higgs at the LHC; this would determine the most appropriate center-of-mass energy of the future LC. There is some hope that by end 2012, the existence of a low mass Higgs will be proven or not [2]. In case there is not such an indication we would have to wait a few years more (the LHC will go through a long shutdown in 2013-2014) before knowing what is the best machine, i.e. an ILC (cold slow machine) or CLIC (warm fast machine). The choice of the machine technology will drastically impact on the design of the detectors and especially on the associated electronics.
Taking these two series of facts into account, the strategy and the perspectives of the R&D on large area Silicon tracking for the future Linear Collider as conducted by the SiLC R&D collaboration [3] over the last decade, are being revisited. This report first describes the present status of the Silicon tracking technology, with the current performances at the LHC experiments and without forgetting the pioneering results achieved by the CDF experiment at the Tevatron. The main upgrade stages that the tracking systems at LHC will undergo over this decade are then presented. The list of main goals to be achieved by the Silicon tracking technology in order to fit with the Linear Collider case are reviewed for both the ILC and the CLIC cases. A few relevant examples then describe how the LHC upgrades and other R&D's or experiments in HEP or different fields are providing useful synergies with the R&D activity for the Linear Collider. This section concludes with a proposed R&D strategy for the years to come including long-term perspectives. The last section briefly mentions, as an example of a new perspective, the development of the Silicon Pixel Tracker (SPT).

# 2 Main changes over the last years and impact for this decade.

The R&D activity for developing the next generation of large area Silicon tracking for the future LC, can be divided in three phases. The first phase goes from 2002-2010, the second phase, covers the present period (2010-2012) and the third period goes from 2013-2020.

The first phase represents the golden years of the R&D for the LC. During this period of time, the LHC experiments were constructing their detectors and thus facing the hard reality challenges for building for instance about 200 $m^2$ of Silicon tracking for CMS. The largest Silicon tracking system ever built until then was the CDF Si tracker covering a total of 7 $m^2$ made of strip sensors. The LC was the unique project proposing to build large area Silicon tracking systems for the two detector concepts. The ILD concept [4] includes a Silicon Envelop [5] fully surrounding the central Time Projection Chamber (TPC) and the SiD concept [7] is based on an all-Silicon tracker a-la-CMS. The CLIC project, launched in 2007-2008 borrowed the two ILD detector concepts with their corresponding Silicon tracking systems. During this first phase a series of impressive advances were achieved on all this technology aspects, i.e. new sensors, new associated Front End Electronics (FEE), mechanical prototyping, alignment technology, full integration studies, detector design and tests. Most of this was summarized in the LoI for the ILD [4] and SiD [6] detector concepts submitted to the IDAG committee in 2010, and in the reports from the SiLC R&D collaboration also part of the EUDET E.U. project [7].

A second phase aiming to the Detector Baseline Design (DBD) is currently underway. This phase covers 2011-2012 and is facing important changes. The LHC experiments have been successfully running now for two years 2010-2011 with relatively high luminosity and 7 TeV c.m. energy. Over these last years, they launched a very aggressive R&D in order to develop, in several stages, a completely new Silicon tracking system that will replace step by step the present one (see next sections). The success of the LHC trackers has further encouraged other experiments or fields to adopt this technology.
In addition, the decision to build the LC machine is waiting for the results on low mass Higgs. This will determined what machine to build: ILC or CLIC? The LC R&D has therefore to cover both cases, each with sometime different issues, in several aspects (see next section). Moreover the financial support for the LC R&D has drastically diminished these last years.
These facts have changed the overall LC R&D scenario. This is particularly true for this technology, because its strong synergy with LHC and its success that made it to be adopted in many other High Energy Physics (HEP) or Astro-particle and Astrophysics experimental fields. Taking into account this overall evolution, we are *revisiting our R&D strategy and studying the new perspectives in order to cope at best with these new times.*

The third phase will start in 2013 and cover the next 10 years, i.e. 2013-2022. Around 2020-2022 the overall upgrade of the LHC experiments for the High Luminosity (HL)-LHC will be achieved. This period is subdivided into two periods: the first period extends from 2013-2015 and is called Phase 1. In this phase, the LHC machine will be upgraded to run at its nominal c.m. energy, i.e. 14 TeV, and reach $10^{34}$ $cm^{-2}$ $s^{-1}$ instantaneous luminosity. The second period from 2015-2020 covers the Phase2 construction for the HL-LHC run at $5x10^{34}$ $cm^{-2}$ $s^{-1}$. These major machine upgrades request important upgrades of all the LHC experiments and especially ATLAS and CMS. They are very demanding in terms of physicists, technical staff and of industrial involvements and contributions on many aspects, without speaking about the cost. For instance two completely new $200m^2$ Silicon trackers will have to be constructed in about 5 to 7 years with very challenging technological requirements of all kinds. The industry will have to provide in this period of time the needed amount of sensors to equip both detectors, same for the associated electronics etc… Thus how to include in such a scenario the R&D for LC followed by the construction of an additional Silicon tracking system of the same dimensions and difficulty, almost contemporarily to the two other ones? *A common sense answer is to exploit at best the synergies between LHC and LC and even with other fields or other HEP experiments R&D's in order to share at best expertise and means.*
In this third phase the LC project may be confronted to the following scenarios. First (most optimistic case) if a light Higgs is discovered by end of 2012 this will provide the green light for building the ILC machine. Secondly, if the preliminary hints on Higgs, shown end 2011, are not confirmed with additional data by end of 2012, the decision on the LC will be delayed by 3 or 4 years, because also of the LHC long shutdown in 2013-2014. Thirdly, if by end 2015 there is no evidence of a light Higgs, CLIC will be the option with an experiment starting around 2025-2030. Thus for the time being, the two options, i.e. the slow and fast LC machine cases must be considered by the detector R&D's.

# 3 New times, new perspectives, revisited R&D strategy and goals

The current status on the large area Silicon trackers R&D is briefly described in this section, stressing the synergy between several fields of application of this technology. The main R&D goals still needed to apply this tracking technology to the future LC machine (cold or warm) are reminded. The undergoing R&D activities for the LHC upgrades or other projects are tackling many of these issues; we outline with a few important examples how to profit at best of this synergy and we revisit/redefine the R&D strategy for the next years to come.

## 3.1 R&D on large area tracking: where do we stand

The current running of the large area trackers at LHC are an ''incredible playground and test facility'' for the development of the next generation of these devices. Firstly it shows the outstanding performances that such tracking systems are able to achieve. It goes sometimes beyond what was anticipated particularly in such harsh environmental conditions. The longtime running (over one decade) of the Silicon tracking system at the CDF experiment brings another interesting piece of information. It teaches us 10 years of operation in an already harsh environment, where the Silicon tracking system did support much higher radiation levels than expected [8]. Moreover the use of this device in the trigger is a pioneering example of such a real time or ''intelligent '' signal processing [9].
These tracking systems currently running in challenging frontier experiments are an invaluable source of lessons on how to develop the next generation Si trackers. We are mentioning some of these main aspects in subsection 3.3.
The current LHC upgrades are requesting a very active R&D on all the aspects of the Silicon tracking technology with requirements that are similar and sometime even more demanding than those for the future LC case. There are however a few main differences between these two cases, namely: including the *power cycling* in the LC case and the strong constraints in *radiation hardness* for the LHC experiments leading to stronger requests on the associated cooling system. Moreover if the LC machine is of the cold type, i.e. a slow machine, the FEE and readout electronics will be relatively slow thus imposing a long shaping time and a different strategy in digitizing and signal processing then the one in the warm case. Moreover, if the LC machine is a fast machine (CLIC), it will be slightly more demanding than the LHC (also a fast cycle machine) in terms of fast Front End Electronics (FEE).
But many R&D features are quite similar and it is indeed mandatory for many reasons: efficiency, industrial interest, cost and available manpower and expertise to develop as much as possible joint efforts between the LC and the LHC R&D activities, and even as we will see with other R&D efforts in other fields or with other HEP projects.

## 3.2 Main R&D goals for the ILC and CLIC large area Silicon trackers

The main parameters that have to be worked out and improved over the next years of R&D are summarized in the Table 1 here below. The star sign (*) added to some items in this Table, indicates the items where the listed goals are the most demanding.
The main parameters to be optimized for the future LC Si trackers as listed in Table 1 are:
*i) The material budget (%X0):* this is an item where a lot of R&D effort is also underway for the LHC upgrades. The thickness of micro-strips sensors is reduced and the edgeless sensors are developed. Even more important is the reduction of the power dissipation of the Front End electronics on the detector and trying to apply the power cycling. It impacts on the cooling system requirements. The cooling system is an important cause of the increase of %X0, together with the cabling and the services and support structures. However, compared to LHC, the LC has lower constraint on cooling because the much lower radiation levels do not require the sensors to be kept to a low and very stable (small gradient) temperature.
ii) *The performances* are in terms of spatial resolution (high granularity), overall coverage, high precision in momentum measurements. The Silicon tracking technology is well suited to these goals. The most demanding conditions will be on the FEE. It mainly impacts on the choice of the appropriate multiplexing factor for the digitizer, the optimization of the Signal over Noise ratio. The use of Deep Sub Micron (DSM) CMOS technology and improved interconnect packaging between sensors and FEE will be instrumental for improving the overall performances of these devices.

iii) *Time stamping and tagging* are important features for these detectors. The most crucial component in this aspect is the FEE, but the choice of the sensors also matters (see 3.3.4).

iv) *Real time processing*: including intelligence in the FEE is another active area of R&D for many different reasons and is important as well for the LC (see 3.3.2). It will indeed strongly impact on all the R&D aspects, i.e. the sensors (direct connection between sensors and FEE), the design of the FEE ASIC and consequently of the overall DAQ architecture. It may as well have related impact on the mechanical architecture and the cabling/services.

v) *Easy to build detector system*: This includes the repeatability and modularity of the overall detector architecture, the lowering of its weight, its easy access and easy to position. These features lead to studying the possibility to have a unique sensor type as elementary component for the overall detector architecture, to optimizing the multiplexing factor of the digitization at the FEE level. Clearly the main way to reach this goal is by optimizing the overall architecture design, the support structures, the distribution/organization of signal cables, the services and the cooling system. Designing and developing ''intelligent'' tracking devices is also there a major asset and R&D goal (see next subsections).

| Parameters/ R&D items | Sensors | FEE | DAQ/ processing | Mechanics/ integration | Cable/cool ing/service |
|---|---|---|---|---|---|
| %/X0 | Thinner sensors | Reduce watt/ch* | Reduce signal cables/fibers | Light structures* | Intelligent cooling/ser vice* |
| Performances:<br>- Granularity<br>- Coverage<br>- Precision<br>- Rad. hard | Optimize<br>50µpitch<br>OK<br>OK<br>OK | Optimize<br>Multiplex<br><br>S/N<br>DSM OK | | Optimize:<br><br>architecture<br>positioning | |
| Time/stamping and tagging | Some impact | Strong impact* | Some impact | Some mech. issues | Possible service issues |
| Real time processing | FEE inte-gration* | Strong design impact* | Strong design impact* | Some mech & integra-tion impact | Possible service issues |
| Easy to build:<br>- repeatability<br>- modularity<br>- weight<br>- easy-access<br>- easy to position | one sensor type/geom | Multi-plexing | Follow FEE basic archit. | High impact* on all these items | High impact* on all these items |

Table 1: Main R&D goals and parameters to be further developed for the LC Si trackers.

### 3.3. Examples of synergy with R&D for LHC or other projects

There are many examples of synergy between the R&D on Silicon tracking for the LC and the R&D underway for the LHC upgrades or other projects. Some among the most important ones are briefly stressed here below.

*3.2.1 The next steps in the sensor R&D technology*

Over these last years, the LHC experiments have studied many possible Silicon strips and pixel technologies including some among the most innovative ones[10]. The development of very hard radiation materials is one of the key-issue for the LHC, not so much for the LC.

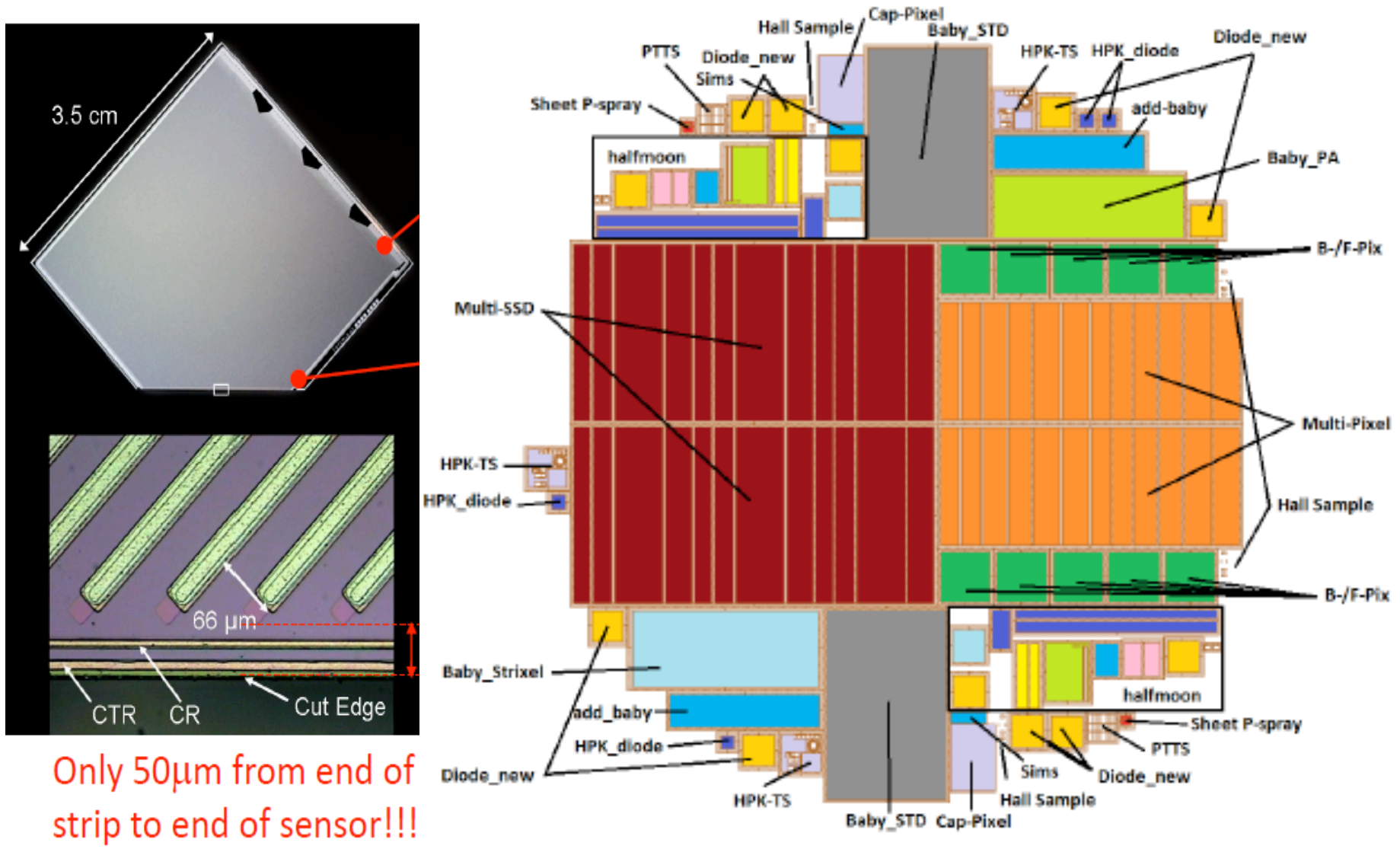

Figure 2 : Photograph of a 3.5x3.5cm2 strip edgeless sensor developed by VTT for TOTEM (left) ; Details of a mask for HPK Silicon detector with many different test structures (right).

But the LHC experiments are working on other sensor issues, studied as well by the SiLC R&D, namely: thinner and/or edgeless sensors (Fig. 2) and inline pitch adapters for strip sensors. Likewise test structures are designed in order to allow the full detector characterization and the optimization of the choice of the sensor material and design (Fig. 2). Advanced short strips (strixels) or pixel technology are explored by LHC and by other projects as the Super B factories. Among them, the 3D pixel technologies are actively pursued and for instance the new B-layer of the ATLAS [11] vertex detector system will include a small part equipped with 3D pixels. Another important aspect of the new sensor technology is the development of new packaging techniques for optimizing the connection between the FEE and the sensor (Fig.3), discussed in the next subsection and with impact on the sensor design.

*3.2.2 Intelligent trackers: the Front End and Far End Electronics challenges*

Intelligent trackers are based on devices with high processing power. This is achieved by embedding "intelligence" thus processing power onto the detector. The on-detector FEE processing is a main R&D goal. It means a FEE readout with high processing capability in Very Deep Sub Micron (VDSM) CMOS technology and optimized sensor-to-FEE connection. Because both the sensor and the associated FEE are based on CMOS technology, having these two parts closely linked is a main R&D goal. It can be achieved by embedding the FEE in the detector substrate and/or through advanced interconnect packaging technologies.
Advanced interconnect packaging solutions are explored for pixels and strip sensors; among the appealing ones, there is the 3D vertical interconnect packaging and the Through Silicon Vias (TSV). Both the SuperB and the LHC experiments (ex one option to build the Pt-layer for the CMS tracking trigger upgrade) are developing this alternative 3D packaging technology for different purposes [12] (Fig. 3).

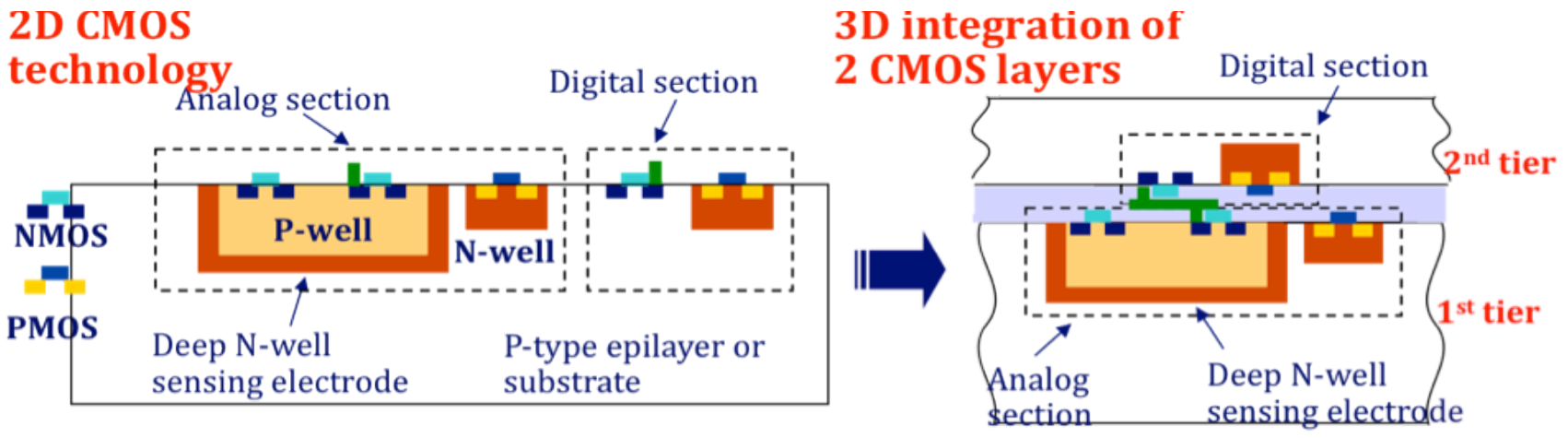

Figure 3: Example of 3D vertical inter-connection design developed for the Super B project.

The SiLC R&D has achieved a pioneer work on both the FEE readout chip and the interconnection sensor-FEE in the case of strip sensors. A mix mode device including pulse-height reconstruction full digital processing and control was developed in 130nm IBM technology [13] and strip sensor prototypes with "*inline pitch adapters*" were designed and built [14]; both features were tested on prototypes at CERN test beams. The LHC experiments are revisiting these ideas and developing them in their new detectors based on strip sensors.
Furthermore, the LHC experiments are developing the signal processing techniques, by including the tracking information in the trigger system of ATLAS and CMS experiments with different approaches [15]. CDF developed a pioneering tracking trigger by including the vertex Silicon detector data in the Level 2 trigger. This real time processing was achieved using associative memories, revolutionary processing devices at that time [16]. The tracking trigger developed for LHC upgrades, is based on a real time processing achieved in very challenging conditions i.e. fast and high data rates requesting massive parallel computing. Thus it has interesting outcomes for the next generation experiments even if not trigger based experiments such as the future LC. Indeed other fields such as Astroparticle or AstroPhysics are adopting these new real time processing technologies both for the front end and the far end (high-level) processing.

### *3.2.3 Intelligent trackers: including intelligence in the mechanical structures*

The large area trackers presently built and running at LHC taught important lessons on the mechanical side as well. First lesson was about decreasing the number of sensor types in order to simplify the sensor production, characterization and mechanical assembly, without mentioning the cost. The ILC detector concepts and especially ILD gave a special attention to this aspect. SiLC R&D has designed the overall Silicon Envelope architecture with only one type of Si strip sensor [4]. This feature is now adopted as well by the LHC upgrades.
The mechanical structures are more and more not just support structures but they include other functions such as cooling, cabling, services, positioning in an optimized way together with a drastic reduction of the material budget. Various approaches are studied for the LHC upgrades or for the Super B factories. The Super B project, develops a novel micro-tubes cooling technology embedded in the Carbon-Fiber Reinforced Polymer (CFRP) support structures [17]. The BELLE II experiment optimized the support structure using the origami technique [18].
This work is instrumental as well for the ILC and CLIC projects. As an example, two alternative support structure designed for ATLAS Phase 2 upgrade based, one on staves the other one on super-modules are shown (Fig.4). These are self-supporting light but robust structures assembled and inserted in the corresponding barrel layers. At the very-end at the detector edge, an independent part hosts the module services, the cooling and signal cabling connections. Because of its location at the detector ends, this strategic part can be replaced rather easily. Figure 4 here below shows the details of this ATLAS design for both cases.

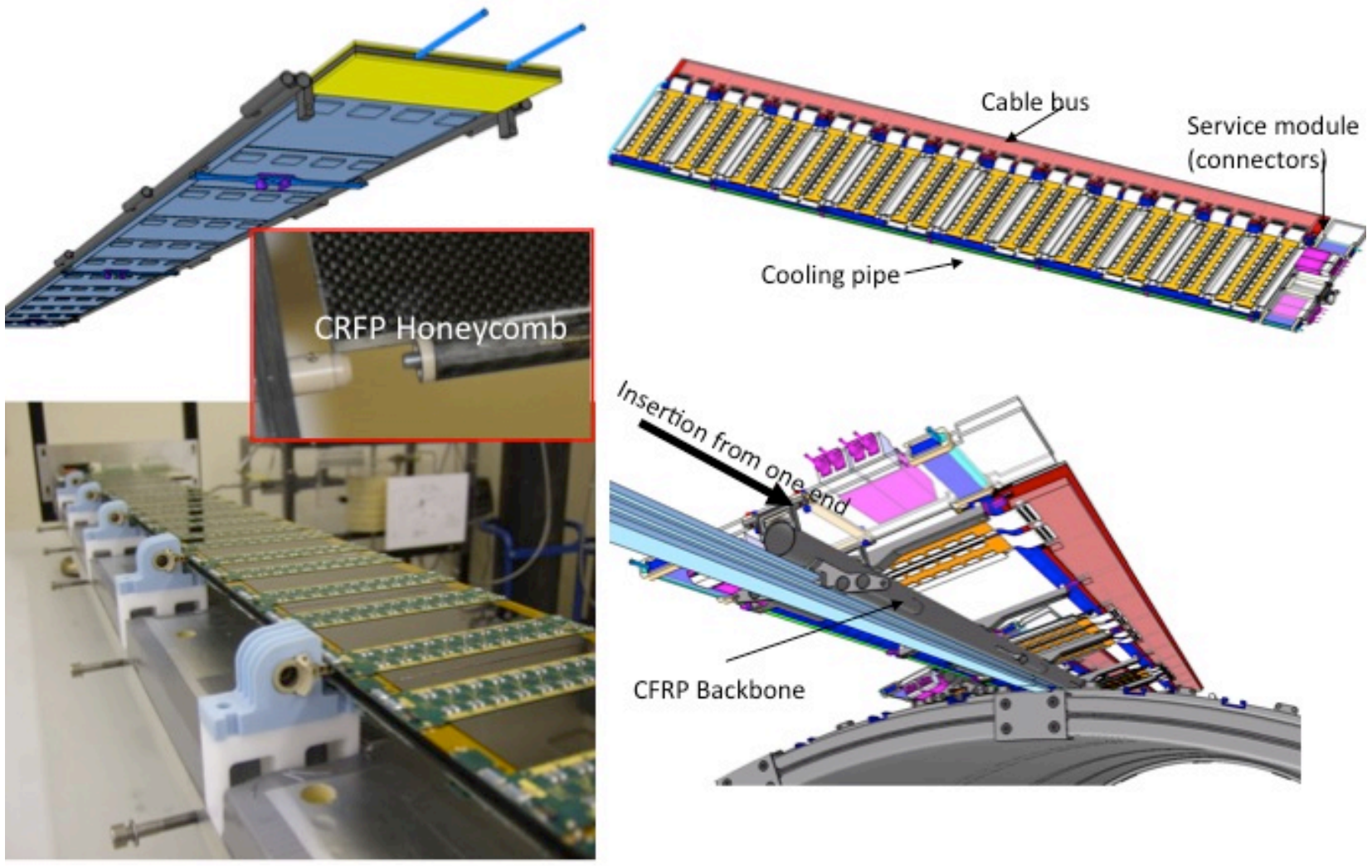

Figure 4: Example of the staves- (left) or super-modules (right) –based supports (ATLAS)

Another important lesson from the LHC commissioning was how well the alignment was accurately achieved with the cosmic ray runs. It showed, that with a precise enough mechanical positioning, sophisticated alignment systems such as the one based on novel semi transparent micro-strip detectors for infrared laser alignment of trackers studied and developed by SiLC, are not may be really needed [19].

### *3.2.4 Intelligent trackers: Time stamping and tagging issues*

The time stamping is another important feature for the LC Silicon tracking and especially demanding for the CLIC option. Indeed ILC, with 300 ns per bunch crossing, only needs about 100 ns time stamping. CLIC, with 0.65ns bunch crossing requires a more stringent time stamping. However because the detector occupancy in both ILC and CLIC will remain relatively low, thus 10 ns time stamping should be sufficient at CLIC.
Implementing a dedicated and refined time stamping device is demanding both on the sensor and associated FEE. It leads to additional power dissipation and increase in material budget. Thus the idea to install a dedicated time stamping layer at the outer part of the tracking system. The additional %X0 can be used as ‘’pre-shower’’ in front of the electromagnetic calorimeter. Another interesting study is reported in this workshop by the ILD detector concept study at CLIC [20]; by combining the TPC and the SET outer Silicon Envelope component with 50μm pitch, a time stamping accuracy of 2.8 bunch crossings is achieved in this simulation study. It shows an asset of the hybrid TPC-Silicon tracking, promoted by the ILD concept.
There are other interesting advanced developments on time tagging in order to get less than 1ns precision. NA62 experiment developed the NA62 Giga-tracker prototype, which is a low mass and sub-nanosecond time resolution silicon pixel detector. By applying a time-over-threshold correction (pixel-by-pixel) using scintillator information, a time resolution of the order of 175ps at 300 V sensor bias is obtained with this prototype [21]. Is this result achievable in a large size detector? Even if such a precision in time is not needed, improving the time stamping and time tagging precision at the nanosecond level may be appealing for CLIC. This is addressed in an interesting way in the SPT proposal (see Section 4).

### 3.3 Revisited R&D strategy and new perspectives for the years to come.

Because of the many changes over the last years mentioned above, the R&D strategy for the Si tracking technology for the LC project must be revisited. Exploiting the synergy especially with the LHC upgrades becomes even more mandatory. It indeed already started in many ways with the involvements of the physicists and the technical staff in the LHC tracker upgrades and related R&D. In some sense it is the same strategy applied by several highly skilled LHC teams when they joined the LC R&D effort in the early 2000. This interchange of expertise and manpower is the most realistic and efficient way for the LC R&D to pursue in these hard times while waiting for the decision to build the future LC. It is an essential feature of the Si tracking technology because so widely applied and it is an important asset. The lack of funding and of people (best ones are taken by the LHC upgrades or other appealing projects) would otherwise dramatically prevent pursuing this activity.
Because the LC schedule is not yet known and may last longer than anticipated it is of the upmost importance to keep an opened eye on new technology and perspective that may seem too adventurous for the present R&D goals for the HL-LHC. The example of the Silicon Pixel Tracker (SPT), briefly mentioned in the next section is one of the appealing ones. In this sense the LC project is a unique training camp to develop long-term interesting R&D.

## 4 Example of a new perspective: the Silicon Pixel Tracker

The Silicon Pixel Tracker (SPT) [22] opens a promising and long-term perspective with important outcomes not only for HEP but also for Astrophysics and astro-particles or medical and other real life applications. This is triggered by the demand in increased granularity (higher and higher precision measurements) together with the striking advances in all the aspects of the semi-conductor technology. Ch. Damerell is developing a novel Silicon Pixel Tracker concept based on the thin monolithic charge-coupled CMOS pixel technology. It was reviewed in this presentation at LCWS11, but we refer to the report by Ch Damerell and K. Stefanov at the VERTEX 2011 Conference for the most complete and up to date description of this new tracking concept [22]. This R&D is perhaps more timely with LC than HL-LHC.

## 5 Concluding remarks

Many of the new ideas put forward by SiLC are now part of the R&D for the upgrade of the LHC Silicon trackers. This was the reverse case when this LC R&D started in the early 2000 and the LHC experts brought their valuable experience. Because the strength and support is now devoted to LHC, it is essential to unite the efforts and LC R&D developers must take part to the R&D LHC activities. This is timely with the LC schedule and it optimizes the needed transfer of knowledge and expertise between these two crucial HEP areas. There are essential communicating vessels between the LHC and LC both in people and expertise especially in this R&D field and this must be exploited by all means and for the benefits of all.

## 6 Acknowledgments

Many thanks to the LCWS11 organizers for the fantastic way they organized this event in such a special place; they ensured the great success of this workshop in not so easy times.